\documentclass[superscriptaddress,
  aps,
  showpacs]{revtex4}
\begin{document}

\title{%
  Decoupling of kinematical time dilation
  and gravitational time dilation\goodbreak
  in particular geometries}

\author{A.\ Radosz}
\affiliation{Institute of Physics, Wroc\l aw University of Technology,
  Wybrze\.ze Wyspia\'nskiego 27, 50-370 Wroc\l aw, Poland}
\author{A.\ T.\ Augousti}
\affiliation{Faculty of Science, Kingston University,
  Kingston, Surrey KT1 2EE UK}
\author{K.\ Ostasiewicz}
\affiliation{Institute of Physics, Wroc\l aw University of Technology,
  Wybrze\.ze Wyspia\'nskiego 27, 50-370 Wroc\l aw, Poland}

\begin{abstract}
Two different forms of time dilation, namely, the kinematical time dilation of
special relativity and gravitational red shift are {\it coupled\/} during
observations of systems moving through a gravitational field. In the
particular situation of free fall in a Schwarzschild geometry these two
effects are decoupled and in consequence the time dilation, as observed by a
distant observer, factorises. Such a factorization is not a universal feature.
We define here a necessary and sufficient criterion for time dilation and
gravitational red-shift decoupling. This property is manifested in a
particular form of the Doppler shift in Schwarzschild geometry.
\end{abstract}


\maketitle

There are two types of time-dilation-like effects: a special relativistic
effect of kinematical origin and a gravitational red-shift (in the case of a
light signal sent by a stationary observer and received by a distant, inertial
observer). In general these two types of time-dilation contributions are
expected to be coupled in the situation where a frame moves in a
gravitational field in an arbitrary way. In this paper we discuss particular
geometries where these time-dilation-contributions are factorizable. We start
by discussing a case set in a Schwarzschild geometry.

Let us consider the case of an in-falling observer in an isotropic, static
gravitational field, i.e. in a Schwarzschild geometry,
\begin{eqnarray}
  {\rm d}\tau^2&=&\left(1-{r_{\rm S}\over r}\right){\rm d}t^2-
                  \left(1-{r_{\rm S}\over r}\right)^{-1}{\rm d}r^2-
                  r^2{\rm d}\theta^2-r^2\sin^2\theta\,{\rm d}\varphi^2
\nonumber\\
          &\equiv&g_{tt}{\rm d}t^2+g_{rr}{\rm d}r^2+
                  g_{\theta\theta}{\rm d}\theta^2+
                  g_{\varphi\varphi}{\rm d}\varphi^2.
\label{Eq1}
\end{eqnarray}
In the case of radial free fall, the velocity components
\[
  \tilde u=(u^t,u^r,0,0),\qquad
  u^t={{\rm d}t\over{\rm d}\tau},\quad u^r={{\rm d}r\over{\rm d}\tau},
\]
are found from an energy conservation:
\begin{eqnarray}
  g_{tt}u^t&=&A,
\label{Eq1a}\\
        u^r&=&\sqrt{A^2-g_{tt}}.
\label{Eq1b}
\end{eqnarray}
The corresponding Killing vector is
\begin{equation}
  \eta^\alpha=\delta^\alpha_t.
\label{Eq1c} 
\end{equation}

The parameter~$A$ is positive and,
\begin{enumerate}
\item[a)]
  $A=1$ corresponds to the case of free fall from infinity,
\item[b)]
  $0<A<1$ corresponds to free fall from a finite distance from the
  gravitational centre,
\item[c)]
  $A>1$ corresponds to the case of an in-falling body, beginning at infinity
  with velocity
  \[
    V_\infty=\sqrt{A^2-1\over A^2}.
  \]
\end{enumerate}

The velocity time-component (\ref{Eq1a}),
\begin{equation}
  u^t={A\over g_{tt}},
\label{Equa5}
\end{equation}
describes the dilation of (coordinate) time~$t$ with respect to the proper
time~$\tau$ measured by the in-falling observer. This expression has a special
meaning: it turns out to be composed of three time-dilation factors of
distinct origins.

To show this let us identify the following observers: apart from an in-falling
observer (IFO), we consider also a distant, inertial observer (IO) (placed at
infinity), and a local observer, static (LSO) with respect to the gravitational
field. The standard gravitational time dilation, the gravitational red-shift,
is described by a factor we term here~$\gamma_{\rm g}$ ({\it g-factor\/}):
\begin{equation}
  \gamma_{\rm g}={1\over\sqrt{g_{tt}}}.
\label{Eq2}
\end{equation}

Right hand side of (\ref{Equa5}) may be decomposed as:
\begin{equation}
  {A\over g_{tt}}=\gamma_{\rm g}{A\over\sqrt{g_{tt}}},
\label{Eq3}
\end{equation}
that provides a residual factor
\begin{equation}
  \gamma_{\rm s}={A\over\sqrt{g_{tt}}}.
\label{Eq3a}
\end{equation}
That factor, which we term the {\it s-factor},
originates from the motion of an in-falling body. As an in-falling object
reaches point~$r$, with a corresponding component of the metric tensor
$g_{tt}(r)$, its velocity, as measured by the LSO, is
\begin{equation}
  v_{\rm IFO}=\sqrt{A^2-g_{tt}(r)\over A^2}.
\label{Eq4}
\end{equation}
In fact, the LSO measures the energy of an object O that passes nearby, by
taking the scalar product of momentum vector of O, $\tilde p_{\rm O}$, and his
own four-velocity~$\tilde U_{\rm LSO}$ (see, e.g., \cite{Cite1,Cite2}):
\[
  E=\tilde p_{\rm O}\circ \tilde U_{\rm LSO}.
\]
The four-velocity $\tilde U_{\rm LSO}$ is a unit time-like vector, defined by
the absolute standard of rest, (\ref{Eq1c})
\begin{equation}
  \tilde U_{\rm LSO}=(\eta^\alpha \eta_\alpha)^{-1/2}\tilde\eta
  ={1\over\sqrt{g_{tt}}}\tilde\eta,
\label{Eq5}
\end{equation}
and the energy is found from
\begin{equation}
  E=g_{\alpha\beta}U_{\rm LSO}^\alpha p_{\rm O}^\beta
  =\sqrt{g_{tt}}\,p_{\rm O}^t.
\label{Eq6a}
\end{equation}

Expressing that energy in the form (see \cite{Cite3})
\begin{equation}
  E={m_{\rm O}\over\sqrt{1-v_{\rm O}^2}},
\label{Eq6b}
\end{equation}
where $m_{\rm O}$ denotes the rest mass of an object~O, one finds its velocity
as measured by the LSO
\begin{equation}
  v_{\rm O}=\sqrt{1-{1\over g_{tt}(u_{\rm O}^t)^2}}.
\label{Eq7}
\end{equation}

Inserting (\ref{Equa5}) into (\ref{Eq7}) one obtains expression~(\ref{Eq4})
for the freely falling object (IFO).

Thus, the corresponding s-factor, turns out to be of kinematical origin,
\begin{equation}
  \gamma_{\rm s}={1\over\sqrt{1-v_{\rm IFO}^2}}={A\over\sqrt{g_{tt}}}
\label{Eq8}
\end{equation}
and factorization of (\ref{Eq3}) corresponds to a decoupling of the two
types of time dilations: the gravitational one and the kinematical one,
\begin{equation}
  u^t=\gamma_{\rm s}\gamma_{\rm g}.
\label{Eq9}
\end{equation}

In the case of a body thrown from infinity towards the gravitational centre,
i.e. $A>1$, this interpretation becomes even more illustrative:
$\gamma_{\rm s}$ factorizes further due to two kinematical contributions. The
first contribution relates to the motion at infinity, for a body being thrown at
infinity, with velocity $v_\infty$,
\begin{equation}
  \gamma_\infty={1\over\sqrt{1-v_\infty^2}}=A
  \iff
  v_\infty^2={A^2-1\over A^2}.
\label{Eq10a}
\end{equation}

The second contribution in (\ref{Eq8}) relates to the case of free fall within
the gravitational field. In fact, for free fall from a state of rest at
infinity one finds the velocity as measured by the LSO (see (\ref{Eq7}); this
is ``escape velocity'' --- cf. \cite{Cite1,Cite3}),
\begin{equation}
  v_r=\sqrt{1-g_{tt}(r)}
\label{Eq10b}
\end{equation}
and the corresponding factor, hereafter termed $\gamma_r$, to be
\begin{equation}
  \gamma_r={1\over\sqrt{1-v_r^2}}={1\over\sqrt{g_{tt}(r)}}.
\label{Eq10c}
\end{equation}

Finally $\gamma_{\rm s}$ in (9) is decomposed into (see (\ref{Eq10a}),
(\ref{Eq10c}))
\[
  \gamma_{\rm s}={A\over\sqrt{g_{tt}}}\equiv \gamma_\infty \gamma_r.
\]

Therefore, in the case of radial free fall in Schwarzschild geometry, one
obtains a factorization of three time-dilation contributions:
\begin{itemize}
\item
  $\gamma_\infty=A$ --- due to the initial velocity (at infinity),
\item
  $\gamma_r=1/\sqrt{g_{tt}(r)}$ --- represents a time-dilation-factor
  due to the work performed by the gravitational field and
\item
  $\gamma_{\rm g}=1/\sqrt{g_{tt}}$ --- an intrinsic red-shift due to the
  gravitational field itself.
\end{itemize}

This result, derived here for a radial free fall, has a deeper sense. In fact,
one
can find that it holds in the case of {\it arbitrary\/} motion within a
Schwarzschild spacetime. Indeed, considering an object~O following arbitrary
trajectory, satisfying:
\begin{equation}
  u_{\rm O}^\alpha\cdot u_{{\rm O}\alpha}=1,
\label{Eq11a}
\end{equation}
one finds its velocity with respect to a static observer from (see (\ref{Eq6a}), (\ref{Eq6b}))
\begin{equation}
  {1\over\sqrt{1-v_{\rm O}^2}}=\sqrt{g_{tt}}\,u_{\rm O}^t
  \equiv \gamma_{\rm s}.
\label{Eq11b}
\end{equation}

Therefore, time dilation factorizes into gravitational and kinematical
contributions in this case:
\begin{equation}
  u_{\rm O}^t=\gamma_{\rm g}\gamma_{\rm s},
\label{Eq11c}
\end{equation}
where the velocity is measured with respect the local
static observer.

The interesting feature is that in the case of a Kerr geometry this result
also holds for the particular situation of a free fall along the {\it axis of
symmetry\/} but does not hold for more general types of fall. Taking into
account the fact of decoupling of the time dilations of different origins,
namely kinematical and gravitational, arising in highly symmetrical
situations, one can ask how general is such an effect? Or in other words: what
is the status of the statement that time dilation factorizes into g- and
s-factors?

To answer this question, one can consider the case of an observer traveling
across an arbitrary static gravitational field where the Killing vector is
given by~(\ref{Eq1c}). The static observer, LSO, whose four-velocity vector is
given by (\ref{Eq5}), measures the velocity of a passing-by body~O (see
(\ref{Eq6a})) as:
\[
  U_{\rm LSO}^\alpha u_{{\rm O}\alpha}=
  {1\over\sqrt{g_{tt}}}g_{t\beta}u_{\rm O}^\beta\equiv
  {1\over\sqrt{1-v_{\rm O}^2}}.
\]

Therefore, for the metric $g_{t\alpha}=g_{tt}\delta_\alpha^t$, in which the
time coordinate is {\it orthogonal\/} to the spatial dimensions, {\it or\/}
for such a {\it fall\/} that $g_{t\beta}u_{\rm O}^\beta=g_{tt}u_{\rm O}^t$,
one can find a factorization of the time dilations as given above:
\[
  u_{\rm O}^t={1\over\sqrt{g_{tt}}}{1\over\sqrt{1-v_{\rm O}^2}}
  \equiv\gamma_{\rm g}\gamma_{\rm s}.
\]

The final conclusion is that in the case of particular symmetries, or for the
particular types of motion, a distant inertial observer would find time
dilation to occur in a factorized form, with a gravitational factor,
$\gamma_{\rm g}=1/\sqrt{g_{tt}}$, and a kinematical factor $\gamma_{\rm s}=
1/\sqrt{1-v_{\rm O}^2}$. The velocity in the last factor corresponds to
the one measured by the local static observer.

What are the possible experimental consequences of this effect? The most
obvious, non-trivial observation, is a Doppler effect. One can find that in
the case of a radial fall in Schwarzschild geometry, light sent by an IO is
received by an IFO as a red-shifted signal
\[
  {\omega_{\rm IFO}\over\omega_{IO}}\equiv
  {1\over\sqrt{g_{tt}}}{\sqrt{1-V_{\rm IFO}^2}\over 1+V_{\rm IFO}},
\]
where the first blue-shift term is of gravitational origin, and the other two
red-shift type factors are of kinematic origin. This result will be
discussed elsewhere \cite{Cite4}.

\section*{Acknowledgements}

This paper was carried out in part under the project PBZ-MIN-008/P03/2003
[Polish Committee of Scientific Research] and one of the Authors, A.R., wishes
to express his gratitude for this support.

\end{document}